\documentclass[12pt,preprint]{aastex}

\usepackage{graphicx}

\shorttitle{Long-wavelength FU Ori emission} \shortauthors{Zhu et al.}

\begin{document}

\title{Long-wavelength excesses of FU Orionis objects: flared outer disks or infalling envelopes?}

\author{Zhaohuan Zhu\altaffilmark{1}, Lee Hartmann\altaffilmark{1}, Nuria Calvet\altaffilmark{1},Jesus Hernandez\altaffilmark{1,2},  Ajay-Kumar Tannirkulam\altaffilmark{1},
Paola D'Alessio\altaffilmark{3}}

\altaffiltext{1}{Dept. of Astronomy, University of Michigan, 500
Church Street, Ann Arbor, MI 48109; zhuzh@umich.edu,
lhartm@umich.edu, ncalvet@umich.edu, hernandj@umich.edu,
atannirk@umich.edu, monnier@umich.edu} \altaffiltext{2}{Centro de
Investigaciones de Astronomia, Apartado Postal 264, Merida 5101-A,
Venezuela} \altaffiltext{3}{Centro de Radioastronomia y Astrofisica,
Universidad Nacional Autonoma de Mexico, 58089 Morelia, Michoacan,
Mexico; p.dalessio@astrosmo.unam.mx}

\newcommand\msun{\rm M_{\odot}}
\newcommand\lsun{\rm L_{\odot}}
\newcommand\rsun{\rm R_{\odot}}
\newcommand\msunyr{\rm M_{\odot}\,yr^{-1}}
\newcommand\be{\begin{equation}}
\newcommand\en{\end{equation}}
\newcommand\cm{\rm cm}
\newcommand\kms{\rm{\, km \, s^{-1}}}
\newcommand\K{\rm K}
\newcommand\etal{{\rm et al}.\ }
\newcommand\sd{\partial}

\begin{abstract}
The mid- to far-infrared emission of the outbursting FU Orionis
objects has been attributed either to a flared outer disk or to an
infalling envelope.  We revisit this issue using detailed radiative
transfer calculations to model the recent, high signal-to-noise data
from the IRS instrument on the {\em Spitzer Space Telescope}.  In
the case of FU Ori, we find that a physically-plausible flared disk
irradiated by the central accretion disk matches the observations.
Building on our previous work, our accretion disk model with outer
disk irradiation by the inner disk reproduces the spectral energy
distribution between $\sim$~4000~\AA~ to $\sim 40 $~$\mu$m. Our
model is consistent with near-infrared interferometry but there are
some inconsistencies with mid-infared interferometric results.
Including the outer disk allows us to refine our estimate of the
outer radius of the outbursting, high mass accretion rate disk in FU
Ori as $\sim$ 0.5 AU, which is a crucial parameter in assessing
theories of the FU Orionis phenomenon.  We are able to place an
upper limit on the mass infall rate of any remnant envelope infall
rate to $\sim 7 \times$10$^{-7}$ M$_{\odot}$yr$^{-1}$ assuming a
centrifugal radius of 200 AU.  The FUor BBW 76 is also well modelled
by a 0.6 AU inner disk and a flared outer disk. However, V1515 Cyg
requires an envelope with an outflow cavity to adequately reproduce
the IRS spectrum.  In contrast with the suggestion by Green et al.,
we do not require a flattened envelope to match the observations;
the inferred cavity shape is qualitatively consistent with typical
protostellar envelopes.  This variety of dusty structures suggests
that the FU Orionis phase can be present at either early or late
stages of protostellar evolution.

\end{abstract}

 \

\section{Introduction}
The FU Orionis systems are a small but remarkable class of variable
young stellar objects (YSOs) which undergo outbursts in optical
light of 5 magnitudes or more \citep{herbig77}, with a F-G
supergiant optical spectra and K-M supergiant near-infrared
(near-IR) spectra dominated by deep CO overtone absorption. FU
Orionis objects (FUors) have been modelled as a high mass accretion
disk around pre-main-sequence stars (Hartmann $\&$ Kenyon 1985,
1987a, 1987b) to explain the infrared color excesses and the
variations of both the spectral type and rotational line width with
wavelength (Hartmann $\&$ Kenyon 1996; Paper I).

\cite{herbig77} argued that at least some FU Orionis outbursts must
be repetitive, and \cite{Lee96} suggested that this could be
explained by infall from an envelope to the disk, replenishing the
disk mass for further outbursts. The infall picture is also
suggested by the presence of scattered light envelopes around FUors,
suggesting that they are objects in early stages of star formation
\citep{herbig77,goodrich87}. In the evolution sequence, FUors may
play significant role in transfering a large amount of mass
($\geq$10$\%$) to the central star, which is even higher than the
mass accumulated in the T Tauri phase \citep{leebook}.

However, not all FUors show the large mid-infrared (mid-IR) excesses
that clearly demand dense infalling envelopes. In particular,
\cite{ALS1987} suggested that FU Ori itself had only a depleted or
low-density envelope. \cite{kenyon91} suggested that a pure flared
disk model could explain FU Ori while an infalling envelope was
needed for V1057 Cyg, but \cite{turner97} proposed that both FU Ori
and V1057 Cyg required flattened envelopes. Finally, taking
advantage of IRS spectra obtained with the Spitzer Space Telescope,
\cite{jgreen06} concluded that the SEDs of V1057 and V1515 Cyg
required envelopes and derived crude models for these objects, while
FU Ori and BBW 76 might be explained with flared disks only. Some
recent studies about the silicate features also show FUors can be
classified as two categories and some objects are evolved (FU Ori
and BBW 76) with only disks left \citep{quanz2007}.

The IRS spectra provide us with the opportunity to perform a much
more detailed SED analysis for FUors. In Paper I, we developed
detailed accretion disk models to study the inner disk of FU Ori and
we derived an inner disk size $\sim$ 1 AU. In this paper, we
re-examine the interpretation of mid-IR excesses, taking advantage
of the IRS spectra, with more detailed radiative transfer models, to
resolve the disk/envelope problem. We will describe the
observational data in \S 2. In \S 3, the method to calculate the
temperature structure of the irradiated surface and the resulting
spectrum is described.  Model results for three FU Orionis objects
(FU Ori, BBW 76, and V1515 Cyg) are presented in \S 5.  Finally, in
\S 6 we discuss some implications of our results.

\section{Optical and infrared data }
\subsection{Photometry and Spectra}
Because FU Orionis objects are significantly variable, it is crucial
to minimize differences in the times of observations at differing
wavelengths. We build on the database assembled in Paper I for FU
Ori. For other FU Orionis objects, we also collected optical
photometry from \cite{jgreen06}. These data were obtained in 2004 at
the Maidanak Observatory and all the data collected in this year for
the same band were averaged considering FUors are variable objects.
Near-IR fluxes are from 2MASS point-source catalog (PSC). V1057 Cyg
was observed in June, 2000; V1515 Cyg was observed in November,
1998; BBW 76 was observed in February 1999. As mentioned in
\cite{jgreen06}, because FU Orionis objects are fading slowly, the
extrapolation from the 2MASS epoch ($\sim$1998) to the IRS epoch
($\sim$2004) is not important; thus, we adopted the 2MASS JHK
photometry without any correction. Mid-infrared fluxes are derived
from the {\em Spitzer} IRS spectra observed in 2004
\citep{jgreen06}.

The spectral energy distributions (SEDs) need to be corrected for
extinction to assess the brightness of the mid-IR excess relative to
the central hot disk, which will determine whether or not a flared
disk can explain the observations. To get the right extinction
correction, we de-reddened the optical and near-IR photometry by a
variety of extinction parameters (A$_{V}$). Then we compared these
de-reddened observations with the steady disk spectra calculated
with our disk model (Paper I), in which disk temperatures peak
around 6000 K. Through this comparison, the extinction parameter
which gave the best fit to the model spectrum was chosen for that
object. Because the disk radial temperature distribution may deviate
from the steady disk and boundary layer emission may be present, we
estimate the uncertainty of our extinction values as $\Delta A_V
\sim$ 0.5.

\subsection{Interferometry}
We use both the near-IR and mid-IR interferometry to test our
modelling of FU Ori.  The near-IR interferometric data are from
\cite{malbet05}, who obtained 287 long-baseline interferometric
observations in the H and K bands from 1998 to 2003, with resolution
of AU scales at the distance of FU Ori. The $(u,v)$-plane coverage
is shown in Fig. 1 in their paper and the averaged square
visibilities are provided in Table 3 of their paper. The mid-IR
interferometric data are from \cite{quanz2006}, who carried
measurements from 8~$\mu$m to 13 $\mu$m at three baselines (44.56m,
86.25m, and 56.74m) between October 31 and November 4 2004 with the
Mid-Infrared Interferometric Instrument (MIDI) at ESO's Very Large
Telescope Interferometer (VLTI) on Paranal, Chile.  The visibilities
are provided in Table 5 of their paper.

\section{Model calculations}

\subsection{Temperature structure of the irradiated surface}
We follow the method of \cite{calvet91} to calculate the temperature
structure of the outer flaring surface irradiated by the inner hot
high $\dot{M}$ disk. In this treatment we have not considered the
outer disk self-heating, which is not important for a moderate
flaring disk, but has a non-negligible effect on a highly flared
envelope, as discussed further in \S 5.2. We have assumed the
flaring surface is very optically thick whether it is the disk
surface or the outflow cavity driven into an opaque envelope.
Milne's reflection effect is included and LTE is assumed. Milne's
reflection effect tells us that, without additional heating sources,
the incident radiation alters the temperature structure of the
low-optical depth layers such that all the energy incident on those
layers is radiated back or, in other words, the net fluxes of those
layers are zero. In the case of LTE, when the source function is
determined by local conditions, the transfer equation is a linear
equation and solutions can be superposed.

Specifically, if $T_1(\tau)$ is the temperature structure of the
standard viscous heating disk with zero incident radiation but net
viscous flux
\begin{equation}
 F_{V}=\frac{3GM\dot{M}}{8\pi R^{3}}\left[1-\left(\frac{R_{*}}{R}\right)^{1/2}\right]
\end{equation}
 and $T_{2}(\tau)$ is
a temperature structure with a given incident radiation but zero net
flux, then the temperature structure of the disk with both of a net
flux $F_{V}$ and a given incident radiation is
$T^{4}=T_{1}^{4}+T_{2}^{4}$. We can solve $T_{2}$ by assuming the
incident radiation from the central star/high $\dot{M}$ disk covers
a well separated frequency range from the disk emergent radiation;
details can be found in \cite{calvet91}. If the incident radiation
consists of a parallel beam carrying energy $E_{0}$(ergs cm$^{-2}$),
incident at an
 angle cos$^{-1}\mu_{0}$ to the normal of the disk surface, a fraction
$\sigma$ of the energy in the beam will be scattered and a fraction
$\alpha$ of the incident beam will be truly absorbed and remitted.
The temperature structure corresponding to an atmosphere that has
$F_{V}$ net flux and incident radiation of flux $E_{0}\mu_{0}$ is
\begin{equation}
T^{4}(\tau_{d})=\frac{3}{4}T_{V}^{4}(\tau_{d}+\frac{2}{3})+\frac{\alpha
E_{0}\mu_{0}}{4 \sigma_{R}}
\times[C_{1}'+C_{2}'e^{-q\tau_{d}/\mu_{0}}+C_{3}'e^{-\beta q
\tau_{d}}] \label{eq:t}
\end{equation}
where
\begin{equation}
C_{1}'=(1+C_{1})(2+\frac{3\mu_{0}}{q})+C_{2}(2+\frac{3}{\beta q})
\end{equation}
\begin{equation}
C_{2}'=\frac{1+C_{1}}{\mu_{0}}(q-\frac{3\mu_{0}^{2}}{q})
\end{equation}
\begin{equation}
C_{3}'=C_{2}\beta(q-\frac{3}{q\beta^{2}})
\end{equation}
and
\begin{equation}
C_{1}=-\frac{3\sigma \mu_{0}^{2}}{1-\beta^{2}\mu_{0}^{2}}
\end{equation}
\begin{equation}
C_{2}=\frac{\sigma(2+3\mu_{0})}{\beta[1+(2\beta/3)](1-\beta^{2}\mu_{0}^{2})}
\end{equation}
where $\alpha=1-\sigma$, $\beta=(3\alpha)^{1/2}$ and
$q=\tau_{s}/\tau_{d}$. $\tau_{s}$ and $\tau_{d}$ are the optical
depths at the stellar and disk frequency. For a viscous heating
disk,
\begin{equation}
\sigma_{R}T_{V}^{4}=\frac{3GM\dot{M}}{8\pi
R^{3}}\left[1-(\frac{R_{*}}{R})^{1/2}\right],
\end{equation} while for an
envelope without viscous heating, $T_{V}=0$.

 We assume a two part model: an
 inner steady accretion disk of high mass accretion rate
 $\dot{M}$ with outer radius $R_{in}$, which is optically thick
 but geometrically thin and flat with height $H_{0}$; and an outer optically thick region, $R\geq
 R_{in}$,
 which is heated by absorption of the light from the inner disk.
  The geometrical details are described in the Appendix A and Fig. \ref{fig:1}.
  The values of the quantities $\mu_{0}$ and $E_{0}$ characterizing the incoming radiation to every outer annulus
 are calculated by integrating the mean intensity arriving from every part of
 the inner disk. These quantities depend on
 both of the inner disk physical condition (eg. the peak temperature,
 disk size) and the geometry of the outer disk surface (Appendix A).

We used the same parametrization to describe the flared surface
geometry of the disk outside $R_{in}$ as \cite{calvet91}. The scale
height of a vertically isothermal disk is
\begin{equation}
H=\frac{c_{s}}{\Omega}
\end{equation}
where $c_{s}$ is the sound speed at the disk midplane and $\Omega$
is the angular velocity. For typical disk temperature distributions,
$T\propto R^{-n}$, where 3/4 $\lesssim$ n $\lesssim$ 1/2. Thus, for
a Keplerian disk, $H\propto R^{\gamma}$, where 9/8 $\lesssim \gamma
\lesssim $ 5/4.

For a very optically thick disk, the absorption height $H_{s}$ where
most of the inner disk radiation is absorbed is a nearly constant
multiple of $H$ \citep{dalessio2001}; thus we parameterize $H_{s}$
as
\begin{equation}
H_{s}=H_{0}(\frac{R}{R_{in}})^{\gamma} \label{eq:surf}
\end{equation}
where $H_{0}$ is the disk height of the inner high $\dot{M}$ disk
and $R_{in}$ is the outer radius of the inner high $\dot{M}$ disk,
which is also the inner boundary of the flared outer surface (Fig.
\ref{fig:1}). Thus, for FU Ori, $R_{in}\sim$0.5-1 AU. We adopt this
form to model all objects. When it turns out that $H_{s}\gg H$, or
$\gamma>5/4$, we conclude that the absorbing surface we are
describing is not that of a flared disk but rather represents the
edge of a cavity in a dusty opaque envelope. Our outer disk (or
envelope) structure joins abruptly to the inner flat thin disk. In
reality we expect a smoother joining between inner and outer
regions, but this depends upon the precise geometry of the inner
disk (height,flaring) which in turn depends upon the detailed
viscosity, etc. Thus, the details of our model near $R_{in}$ (and
the precise values of $R_{in}$ and $H_{0}$) are uncertain, while the
structure at $R>R_{in}$ is more robustly modeled.

The gas opacity used in this model is the same as we used in Paper
I. We improve our dust opacity by using the prescription in
\cite{dalessio2001}. For each dust component, the grain size
distribution is given by a power law of the grain radius,
$n(a)=n_{0}a^{-p}$, between a minimum and maximum radius. We choose
$a_{min}$=0.005 $\mu$m, $a_{max}$=1 $\mu$m, and $p$=3.5. The
detailed ingredients of the dust are shown in Table 1. With this
monochromatic dust opacity and gas opacity from Paper I, we
calculate the $\alpha$ and $q$ parameters of Eq. (\ref{eq:t}). For
$\alpha$, we weight the monochromatic $\alpha_{\nu}$ by the Planck
function at the stellar temperature. For $q$, we calculate the
Rosseland mean opacity weighted by the planck function at the
stellar temperature and the disk temperature separately, and then
divide them. We also use the same opacity to calculate the emergent
spectra and disk images. Although the dust ingredients may vary for
different objects, we use the ingredients in Table \ref{tab1} for
all, considering that larger uncertainties caused by the invalidity
of the plane parallel assumption, as discussed below.

\subsection{Emergent spectra and disk images}
Given $\alpha$, $q$, $\mu_{0}$ and E$_{0}$, we can calculate the
temperature structure of every outer irradiated annulus using Eq.
(\ref{eq:t}).  The emergent intensity of any annulus at any outgoing
direction then can be derived by solving the radiative transfer
equation. Our calculation of the emergent spectrum does not include
scattering. Scattering is unlikely to be important in the
mid-infrared but may have an effect in the near-infrared; we will
consider a detailed scattering treatment in a future paper.

When the intensity of every annulus, every outgoing direction and
every wavelength is derived, the total flux and the image of the
flared surface for any inclination angle can be derived, for given
flared surface geometry.
 The total flux can be compared with the IRS spectra while
the image can be used to do interferometric tests. The details are
shown in Appendix B.

\section{Results}
\subsection{FU Ori}

Fig. \ref{fig:fuorispe} shows the final model spectra for FU Ori
compared with the observed SED. The parameters are in Table 2. The
inner disk model is that of Paper I, which shows absorption features
because it is internally heated by viscous dissipation. The outer
flared disk exhibits
 10 and 18 $\mu$m silicate emission features due to
the strong irradiation by the inner high $\dot{M}$ disk, which
produces a temperature inversion at its surface.  The model does not
fit the 18 $\mu$m feature very well, suggesting that either the size
distribution we are using for the dust may vary with radius or the
temperature structure of the outer region is not accurate.

The adopted surface has the parameters $H_{s} = 0.19
(R/R_{in})^{1.125}$~AU, where $R_{in}$=0.58 AU (see Fig.
\ref{fig:fuorisurf}). To investigate whether this surface is
plausibly that of a flared disk, we computed the approximate local
vertical scale height $H = c_s/\Omega$, assuming that the central
mass is 0.3 M$_{\odot}$ (Paper I) and the temperature at the
mid-plane is equal to the temperature at $\tau_{d}$=10 (i.e., the
disk interior temperature distribution is roughly isothermal.).  The
results indicate that our adopted surface lies roughly 3 scale
heights above the midplane (Fig. \ref{fig:fuorisurf}), comparable to
the dusty disk scale heights estimated for typical T Tauri stars
\citep{kenyon87,dalessio1998,dalessio2001}. Thus, a
physically-plausible flared disk can reproduce the mid-IR flux of FU
Ori. The temperature at $\tau_{d}=2/3$ is shown in Fig.
\ref{fig:surftemp}. Because the disk is isothermal when
$\tau_{d}>\sim$$1$, the midplane temperature is almost the same as
Fig. \ref{fig:surftemp}.

In Paper I, without considering the flux from the flared outer
surface, we derived an inner high $\dot{M}$ disk size of $\sim$ 1
AU, constrained by the mid-IR IRS spectrum.  The flared outer disk
contributes some emission shortward of 10 $\mu$m, necessitating a
reduction of the outer radius of the inner outbursting disk from 1
AU to $\sim$ 0.58 AU. This reduction of the outer radius does not
significantly affect the conclusions of Paper I; in particular, the
model of Bell \& Lin (1994) still cannot reproduce the observations.

We next compare model results with near-IR and mid-IR
interferometry. The images at the H and K bands are calculated by
assuming the above flared surface geometry and all the parameters
flow strictly from the SED fitting. To compensate for the lack of
scattering in our models, we add an extra component of emission
corresponding to an ``effective albedo'' of about 30\% in H and K;
compared with the case without scattering, the scattering model
decreases the square visibilities by only 5\%. We adopt the 55
degree inclination angle estimated by \cite{malbet05}. We further
adopt a -30 degree position angle for the disk, which is roughly
perpendicular to the axis of large-scale reflection nebulosity seen
in optical imaging \citep{Lee96}, although due to the large error
bars in the observed visibilities, the
choice of the position angle has little effect.

With the image as shown in Fig. \ref{fig:image}, the resulting model
visibilities are compared with the observations in Fig.
\ref{fig:HKvis}. Within the relatively large observational errors,
our FU Ori model is consistent with the near-IR interferometry.
 Considering
90\% of the flux at H and K bands is from the inner disk, the
near-IR interferometry only tests the inner disk modeling and longer
wavelength data are needed to test the outer disk modeling. We
compute the mid-infrared visibilities, and compare with the
VLTI-MIDI data from \cite{quanz2006}; The comparison is shown in
Fig. \ref{fig:midi}. While the model predictions are in reasonable
agreement with the observations at the shortest wavelengths, the
predicted visibilities are considerably lower than the observations
at long wavelengths. Although self-heating is not important in SED
fitting because only 17$\%$ of the inner disk radiation intercepted
by the outer disk within 10 AU by Eq. 7 in \cite{jgreen06},
self-heating may change the visibility a little bit. However, it is
not apparent to tell whether the visibility will increase or
decrease, because the inclusion of the self-heating will decrease
the flaring slightly, and thus decrease the amount of light
intercepted from the inner disk, to keep the same SED.

 The
comparison with the mid-infrared observation might be improved if
the disk has a higher inclination angle than estimated by Malbet et
al. With a bigger inclination angle, the disk would be less resolved
(higher visibility) for the 44.56 m baseline which lies on the short
axes of the disk image.
 The visibilities for other two baselines
would also increase, though not as much as for the 44.56 m baseline.
However, a bigger inclination angle may also increase the
visibilities at shorter wavelengths which may contradict with the
observed visibilities around 8 $\mu$m. Unfortunately we can not test
this by our model because the light from the inner disk has to
travel through the outer disk to reach us at higher inclination
angle and this needs a more complicated radiative transfer treatment
than adopted in this paper.

Another possibility is that there might be an abrupt change in the
disk absorption height at the radius around $R_{in}$ where dust
condenses, due to the
 much larger opacity of dust than gas; this might produce
a radially-sharp mid-infrared emission feature in the disk where the
dust begins to form at T$\sim$1500 K, by intercepting more
irradiated flux from the inner disk. Finally, there might simply be
narrower structures than we can model, such as non-axisymmetric
spiral arms due to gravitational instability in the outer disk
(Paper I).

Overall, the spectrum fitting and the near-infrared interferometry
support our flared disk model.  Better uv plane coverage in mid-infrared
interferometry would help to improve our understanding of
the outer disk structure.

\subsection{BBW 76}
We fit the SED of BBW 76 following the same procedure as used for FU
Ori (Fig. \ref{fig:BBWspe}). Given the similarity of the large scale
nebulosity to that of FU Ori, we adopt a similar inclination angle
of 50 degrees. 
Because BBW 76 has the same rotational velocity broadening as FU Ori
(Reipurth et al. 2002), we choose the same central star mass 0.3
M$_{\odot}$. The parameters of the best fit model are displayed in
Table 2. The flared surface (solid curve in Fig.
\ref{fig:fuorisurf}) also lies close to 3 scale heights, and is
therefore consistent with a flared disk structure. With the same
relative strength of the IRS spectrum with respect to the optical
flux and the same IRS spectrum shape as FU Ori, we derived a similar
outer radius of the inner high $\dot{M}$ disk $\sim$0.6 AU. However,
the outer disk extends further than in FU Ori, to $\sim$ 200 AU, due
to the slightly higher 18 $\mu$m excess of BBW 76.

\subsection{V1515 Cyg}

The nebulosity seen on large scales suggests that we view V1515 Cyg
nearly pole-on, looking right down an outflow cavity (Goodrich
1987). For simplicity we assume an inclination angle of 0; small
departures from this will have little effect on our modelling.
Because of the low inclination, it is difficult to constrain the
central star mass very well (Kenyon, Hartmann, \& Hewett 1988). We
therefore assume the central star is like FU Ori with a mass of 0.3
M$_{\odot}$. In contrast to FU Ori and BBW 76, the absorption height
of the surface required to fit the SED of V1515 Cyg is highly flared
(Fig. \ref{fig:v1515spe}). This surface lies at about 6 scale
heights (Fig. \ref{fig:fuorisurf}) with the assumption that the
central star mass is 0.3 M$_{\odot}$; this value is higher than a
physically plausible height for a disk. Instead, the large dust
surface height and relatively small opening angle (H/R $>\sim$ 1)
strongly suggest that this surface defines an outflow-driven cavity
in an infalling envelope, as suggested by Kenyon \& Hartmann (1991)
and Green et al. (2006).

Though this surface may extend to a very large radius, most of the
IRS mid-IR flux comes from small radii, eg. 80 $\%$ of 10 $\mu$m
flux and 50 $\%$ of 20 $\mu$m flux come within 10 AU. This can be
roughly estimated from Fig. \ref{fig:surftemp}. By wien's law, 100 K
blackbody has its peak flux at 20 $\mu$m and from Fig.
\ref{fig:surftemp} 100 K corresponds to 10 AU.

Our assumed dust properties fit the silicate emission features of FU
Ori very well, but in V1515 Cyg the model continuum is a bit too
steeply declining longward of 30 $\mu$m; in addition, the model
predicts slightly stronger silicate features than observed (Fig.
\ref{fig:v1515spe}). These discrepancies might be reduced if we had
adopted larger dust grain sizes which can make the silicate features
wider and shallower. However, it could also be that our assumption
of plane-parallel geometry is not correct for the outer envelope. As
Fig. \ref{fig:surftemp} shows, although the fluxes at $\lambda <10
\mu m$ (characteristic of a black body temperature $<$ 300 K) are
mainly from the surface within 10 AU, the long wavelength fluxes may
come from a much larger radius from 10 AU to 100 AU. Because
infalling envelopes are much less optically thick than disks at
these radii, long wavelength photons may travel further into the
envelope, invalidating the plane-parallel assumption.

\section{Discussion}
\subsection{Flared Disks}
The inner disks of the FUors are internally-heated accretion disks,
which produce absorption features. However, the silicate emission
features must arise in a dusty region that is externally heated to
produce a temperature inversion in the surface of optically-thick
dusty disks and envelopes. In the cases of FU Ori and BBW 76, we
find that the IRS SED can be explained with physically plausible
flared disks.

For the flared disks of FU Ori and BBW 76 we derived $H_{s}/R \sim
0.5$ which is larger than the value of 0.2 crudely estimated by
\cite{jgreen06} for FU Ori.  The main reason for the discrepancy is
that we adopt a smaller $A_{V}$ than \cite{jgreen06}, which makes
the inner disk fainter and thus requires a more flared outer disk to
absorb and reemit the same luminosity (the long-wavelength spectrum
is relatively unaffected at these low values of extinctions).  If
anything our extinction estimate is likely to be low, which means
that we may have overestimated the required flaring, and thus makes
it even more plausible that the long-wavelength emission comes from
the outer disks.

We note that, in estimating the number of scale heights for the
disk, we have assumed that the disk has been able to relax to
vertical hydrostatic equilibrium after the outburst. Before the
outburst, we expect the outer disk to be cooler and thus thinner.
After the outburst, the outer disk is heated by the inner hot disk
and gets puffed up. The time scale for the disk to adjust its
vertical hydrostatic equilibrium to the extra irradiation is
$H/c_{s}$, where $H$ is the disk scale height and $c_{s}$ is the
sound speed. Since $H = c_{s}/\Omega$, $H/c_{s} =1/\Omega$ or $P/2
\pi$, where $P$ is the orbital period. Thus,
\begin{equation}
\frac{H}{C_{s}}=\frac{1}{2\pi}\frac{(R/1AU)^{3/2}}{
(M/M_{\odot})^{1/2}} yr
\end{equation}
As the outburst of FU Ori has been proceeding for over 70 years now,
our assumption of hydrostatic equilibrium is good for the inner
parts of the flared disk, but may be less appropriate for the outer
disk; beyond 50 AU for an assumed 0.3 M$_{\odot}$ central star
(Paper I), the hydrostatic equilibrium time starts to become longer
(100 yrs) than the length of the current outburst.

While it is not necessary to include disk accretion energy in the
outer disk to fit the spectrum reasonably well, this does not mean
that the disk is not accreting; the local viscous energy release is
simply much less than the irradiation heating. We find that outer
disk accretion rates greater than $\sim 1/4$ of the inner disk
accretion rate yield silicate features shallower than observed in
the IRS spectra. Thus, the data suggest that the outer disk
accretion rates in BBW 76 and FU Ori are significantly smaller than
the inner disk accretion rates. This is consistent with models in
which material piles up and then accretes in bursts (e.g., Kenyon
$\&$ Hartmann 1991; Armitage et al. 2001).

\subsection{Envelopes }

The IRS SED of V1515 Cyg cannot be explained with a reasonably
flared disk.  A much more plausible explanation is that we are
observing emission from the surface of a cavity in a dusty,
infalling envelope. \cite{jgreen06} also inferred an envelope around
V1515 Cyg, and estimated a maximum solid angle of the outer envelope
of $H_{s}/R \sim 0.37$ based on 14\% of the inner disk radiation
being intercepted and reradiated by the envelope. However, their
estimate was a lower limit because they did not include emission at
longer wavelengths than observed by the IRS.


In our radiative transfer model, 80$\%$ of the 10 $\mu$m flux and
50\% of the 20 $\mu$m flux arises from within 10 AU.  At this radius
the height of the envelope surface is about 10 AU (Fig.
\ref{fig:fuorisurf}), so that the fraction of solid angle covered at
this point is about 70\%. This is roughly twice the solid angle
estimated by  $H_{s}/R \sim 0.37$ from Green et al. , but the
assumed geometries are different. Our cavity structure (Fig.
\ref{fig:fuorispe}) is much more consistent with typical outflow
cavity structures \citep{Dougados2000} and our radiative transfer
methods are more robust. We estimate that self-heating of the outer
flared disks, which we ignore, is a relatively small effect because
the outer disk only intercepts a small fraction of the inner disk
flux. This is not true of the envelope for V1515 Cyg, and so our
results for this object can only be suggestive rather than
conclusive. Our main point is that a flared disk model cannot
reproduce the observations of V1515 Cyg and V1057 Cyg. We are
currently exploring models o these two objects with two-dimensional
radiative transfer and will report results in a future paper.

We have also begun to investigate V1057 Cyg. Our preliminary model
is very similar to that of V1515 Cyg, with a highly flared envelope.
However, when the object is not pole-on or a flat disk our method is
questionable (i.e., maybe we are looking along the edge of the
envelope, or even through some of the envelope), thus we will
revisit V1057 Cyg using a more detailed radiative transfer method in
a subsequent paper.

The near-IR Keck interferometer \citep{Millan2006} also found a
difference between the near-IR sizes for V1515 Cyg and V1057 Cyg
compared with FU Ori. The low near-IR visibilities suggest V1515 Cyg
and V1057 Cyg are more resolved than FU Ori, which suggests dense
envelops may be present for these objects. This is an additional
support for the envelope model, but in future the detailed modeling
is needed to fit these interferometry data.
\subsection{Do FU Ori and BBW 76 have envelopes? }

As both FU Ori and BBW 76 have large-scale reflection nebulosities,
it suggests that they have remnant infalling envelopes, if
sufficiently optically thin. We can estimate an upper limit to the
amount of possible envelope material in FU Ori and BBW 76 from a
limit on the visual extinction. We use the rotating collapse model
of Terebey, Shu, \& Cassen (1984). When $r \gg r_{c}$ ,where $r_{c}$
is the centrifugal radius,
\begin{equation}
\rho\sim\frac{\dot{M}}{4\pi(2GM)^{1/2}}r^{-3/2}\,;
\end{equation}
when $r \ll r_{c}$,
\begin{equation}
\rho\sim\frac{\dot{M}}{8\pi
r_{c}(GM)^{1/2}}(1+\cos\theta)^{-1/2}r^{-1/2}
\end{equation}
Integrating the above equations from 0 to $\infty$ and then
multiplying by the visual dust opacity $\kappa_{V}$ at the V band,
\begin{equation}
\tau_{V}=\int_{0}^{\infty}\kappa_{V}\rho dr
=\frac{\dot{M}\kappa_{V}}{4\pi(GMr_{c})^{1/2}}((1+\cos(\theta))^{-1/2}+2^{1/2})
\end{equation}
With $\theta = 60^{o}$, $M=0.3 M_{\odot}$, $\kappa_{V}$=100
cm$^{2}$/gr, and $\tau_{V}<$2 constrained by A$_{V}$$\sim$2, we find
\begin{equation}
\frac{\dot{M}}{r_{c}^{1/2}}<\frac{7\times10^{-7}M_{\odot}yr^{-1}}{(200
AU)^{1/2}}
\end{equation}
If $r_{c}$=200 AU, we thus estimate an upper limit of $\dot{M}
\lesssim 7 \times 10^{-7} M_{\odot}$yr$^{-1}$, which is considerably
smaller than typical Class I infall rates of $2 - 10 \times 10^{-6}
M_{\odot}$yr$^{-1}$.

\cite{ALS1987} also suggested that FU Ori has a highly depleted
dusty envelope, with a density depletion factor 0.01 with respect to
the density derived by their standard infall envelope model. Though
they didn't give the mass infall rate for the depleted envelope
models, we can estimate their mass infall rate is
$\sim$10$^{-7}$M$_{\odot}$/yr by multiplying the depletion factor
0.01 by their standard mass infall rate 10$^{-5}$M$_{\odot}$/yr.
Because they didn't consider the flared outer disk which is the main
contributor at IRS range, this infall rate is an upper limit.
Submillimeter observations barely resolve FU Ori and BBW 76
\citep{2001ApJS..134..115S}, which also implies highly evolved or
absent envelopes.

\subsection{Dust properties}

The 18 $\mu$m features in the SEDs are not well fit, indicating that
we may need to use a different grain opacity or a more detailed
radiative transfer method. However, we can still study the dust
properties from the 10 $\mu$m silicate emission features. The dust,
with a power-law distribution of the grain radius and a maximum
radius 1 $\mu$m, is larger than the typical interstellar medium dust
with a maximum grain radius 0.25 $\mu$m \citep{dalessio2001}.
Although this dust composition differs from the one used by
\cite{quanz2007}, where they use a mixture of grains with three
different sizes (0.1,1.5 and 6 $\mu$m), it also suggests that grain
growth has taken place in the disk. Our results confirm
\cite{quanz2007} that a significantly higher fraction of
$\sim$1$\mu$m grains is required. BBW 76 is similar to FU Ori while
V1515 Cyg seems to require grains growing to even larger sizes.

In Paper I, we suggested that dust might be depleted in the inner
disk, based on the observed water vapor absorption features at $8
\mu$m. However, if we deplete the dust in the outer disk of FU Ori
and BBW 76, we predict a strong water vapor emission at 8 $\mu$m,
which is not observed. Thus, there may not be significant depletion
in the outer disk, which may be reasonable given the youth of the
system and the likelihood that grain growth/settling timescales are
longer at larger radii \citep{weiden1997}.  However, this is not a
firm conclusion, as the abundance of water vapor is poorly
understood and the region of the disk where most of the 8 $\mu m$
flux comes from is at the region where we join the inner and outer
disk solutions, and thus is uncertain.

\subsection{Evolutionary states and FU Ori classification}

The (meager) event statistics suggest that there must be repeated FU
Orionis outbursts in at least some young stars (Herbig 1977;
Hartmann \& Kenyon 1996).  Kenyon \& Hartmann (1991) proposed that
continued infall was needed to replenish the material accreted onto
the central star during an outburst. While we are unable to
determine the infall rate for V1515 Cyg from our particular
modelling, previous estimates suggest that sufficient matter may be
falling in to make repeat outbursts possible in V1515 Cyg and in
V1057 Cyg (Kenyon \& Hartmann 1991; Green \etal 2006).

In contrast, we have shown that FU Ori and BBW 76 have little or no infall from
remnant envelopes.  The low upper limits of infall in these objects
suggest that continuing infall may not be important in continuing outbursts;
any subsequent outbursts may have to rely on material already present in these
disks.  Thus, we suggest that FU Ori and BBW 76 are nearing the end of their
outburst lives, while V1515 Cyg and V1057 Cyg have a much higher probability
of continuing their activity.

The significant difference between FU Ori and BBW 76 on the one
hand, and V1515 Cyg and V1057 Cyg on the other, is also supported by
submillimeter observations. FU Ori and BBW 76 are very faint and
hardly resolved at submilimeter wavelengths
\citep{2001ApJS..134..115S}, suggesting that they have more depleted
envelopes than other FU Orionis objects.

Overall, the variety of the FU Orionis
envelopes suggests that FU Orionis phenomenon is observed at
different stages of the protostars and every YSO may experience
several FU Orionis outbursts from the early to the late stages of protostellar
infall.

In this regard, young stellar objects (YSO) have been classified as
two categories based on their spectral index $n\equiv d$ log($\nu
F_{\nu})/d$ log $\nu$ in the near-infrared and mid-infrared
\citep{ALS1987}. $n\leq 0$ represents the protostars embedded in
infalling envelopes (Class I) and $n>0$ represents the T Tauri stars
without envelopes (Class II). This classification helps us to
understand the evolutionary sequence of YSO, from protostars to
pre-main-sequence stars. As we discussed above, FU Orions objects
are in a stage between Class I and Class II. Thus, it is also
helpful to classify FU Orions objects in two categories
\citep{quanz2007}. However because of the strong near-IR flux
produced by the high $\dot{M}$ disk, $n$ is always positive in the
near-IR and mid-IR (1-10 $\mu$m) range. Thus, based on the spectra
and envelope structures of FU Ori, BBW 76 and V1057 Cyg, we suggest
a criterion to classify FU Orionis objects by using mid-IR spectral
index $n\equiv d$ log($\nu F\nu)/d$ log $\nu$ from 14 to 40 $\mu$m,
where most of the flux is contributed by the outer region beyond the
inner disk. If $n$ is negative or close to zero, this FU Orionis
object has a circumstellar envelope which produces a flat mid-IR
spectrum (eg: V1057 Cyg and V1515 Cyg). If $n$ is positive it may
have a remnant envelope far away and the mid-IR flux is contributed
by a flared outer disk (eg: FU Ori and BBW 76).

This classification method is different from the method used by
\cite{quanz2007} where they used the 10 $\mu$m silicate feature to
classify FU Orionis objects. They suggest that the objects with
absorption features are in Category I, embedded in circumstellar
envelopes, while the objects with emission features are in Category
II, which are more evolved and show properties of Class II sources.
However, as discussed above, some FU Orionis objects embedded in
envelopes have cavities (eg. V1057 Cyg and V1515 Cyg) with moderate
opening angles. Thus, if we view these objects pole on, we will
still detect the emission features and would classify them as
Category I by the emission silicate feature method.

Another benefit of using a  mid-IR index is that the strength of 10
$\mu$m silicate features are dependent on the inclination angle. In
contrast, for continuum emission at 14-40 $\mu$m is not affected
significantly by the inclination and envelope extinction, since the
opacity at 14 $\mu$m is close to the opacity at 30 $\mu$m, and 5
times smaller than the opacity at 9 $\mu$m for a$_{max}$=1 $\mu$m
olivine grains. Even for a highly embedded FU Orionis objects,
A$_{V}$=10 (A$_{silicate}$=0.7), A$_{14\mu m}$ is only $\sim$ 0.15,
in which case an extinction correction is not necessary. With the
mid-IR index method, we get the same classification for the objects
listed in \cite{quanz2007} except that V1057 Cyg, V1515 Cyg, and V
1647 Ori are identified in Category I. In this paper, we suggest
that V1057 Cyg, V1515 Cyg have normal envelopes, thus they are
indeed in Category I. V1647 Ori object is a special FUor with both
large extinctions, A$_{V}\simeq$11 and silicate emission features.
\cite{quanz2007} suggests there maybe a foreground cloud between
this object and us, which makes it hard to classify and it needs
detailed modelling.

\section{Conclusions}
With the latest IRS spectra and a detailed radiative transfer model
including irradiation, we revisited the envelope/flared disk problem
of FUors. We confirmed the results of \cite{jgreen06} that FU Ori
and BBW 76 have flared outer disks while V1515 Cyg has a
circumstellar envelope. However, instead of a flattened envelope and
a large opening angle \citep{jgreen06}, we derived a more reasonable
cavity structure.

For FU Ori, near-infrared interferometry also supports our model,
while mid-infrared interferometry does not fit perfectly and needs
further modelling. Including the contribution from the outer flared
disk, we gave a tighter constraint on the outer radius of the inner
disk $\sim$0.5 AU, which is essential to test outburst theories
(Paper I).

We also suggest using a spectral index $n$ from 14 $\mu$m to 40
$\mu$m to classify FUors into early (Category I) and late (Category
II), which is less sensitive to the viewing angle. The variety of FU
Orionis envelopes suggests that FU Orionis phenomenon is observed at
different stages of the protostars, and every YSO may experience
several FU Orionis outbursts from very early to late stages of the
infalling envelope.

This research was supported in part by the University of Michigan.
In addition, this work was based in part on observations made with
the Spitzer Space Telescope, which is operated by the Jet Propulsion
Laboratory, California Institute of Technology under a contract with
NASA. We gratefully acknowledges John D. Monnier who drew our
attention to the interferometry and gave us many helpful advices.


\appendix{}
\section{Flared surface geometry}
 The inner disk and outer flared surface are
shown in Figure 1. For
 point A, located at distance $R_{A}$ from the center, the
quantities $\mu_{0}$ and $E_{0}$ are derived as follows. We define
several vectors as shown in the Figure 1; the normal vector to the
flaring surface at A is:
\begin{equation}
\hat{n}=-sin\alpha\,\hat{y}+cos\alpha\, \hat{z}
\end{equation}
The vector from the shaded area at R$_{i}$ in the inner disk to
point A at the flaring outer disk is:
\begin{equation}
\vec{r}=R_{i}sin\phi\hat{x}+(R_{A}-R_{i}cos\phi) \hat{y}+(H_{A}-H_{0})\hat{z}
\end{equation}
where
\begin{equation}
 \hat{r}=\frac{\vec{r}}{|\vec{r}|}
\end{equation}

The solid angle $d\omega$ extended by S with respect to point A is (
S is the area of the tiny shaded region at R$_{i}$ in figure 1.)
\begin{equation}
d\omega=\frac{\vec{S}\cdot\hat{r}}{|\vec{r}|^{2}}
\end{equation}
where
\begin{equation}
\vec{S}=S\hat{z}
\end{equation}
Thus $E_{in}^{i}$, the amount of energy irradiated from the inner
disk annulus R$_{i}$ to unit area normal to $\vec{r}$ at point A ,
is $I_{i}\Delta\omega$ integrated over all the area of annulus i.
Thus,
\begin{equation}
E_{in}^{i}=\int_{annulus R_{i}} I_{i} d\omega=\int_{0}^{2\pi}
I_{i}\frac{\hat{z}\cdot\hat{r}}{|\vec{r}|^{2}} R_{i} \Delta R d\phi
=I_{i}\frac{2\pi R_{i} sin\gamma \Delta R
}{R_{A}^{2}+(H_{A}-H_{0})^{2}}
\end{equation}
Because of limb darkening, $I_{i}$ from the inner disk annulus i
equals to $I_{i}^{0}\times\frac{3}{5}(\frac{2}{3}+\mu)$, where
$I_{i}^{0}$ is the intensity at radius R$_{i}$ towards z direction
(direction perpendicular to the inner disk surface) and $\mu$ is the
cosine of the angle between $\vec{r}$ and z direction; thus we use
$sin\gamma$ to approximate $\mu$.

 Then, $E_{in}^{i}\mu_{i}$, the amount of energy irradiated from the inner
disk annulus R$_{i}$ to unit area at the flaring surface point A
(notice it is different from the above unit area normal to $\vec{r}$
at point A), is
\begin{equation}
E_{in}^{i}\mu_{i}=\int_{annulus R_{i}} I_{i}\hat{n}\cdot\hat{r}
d\omega
=\int_{0}^{2\pi}I_{i}\frac{\hat{z}\cdot\hat{r}}{|\vec{r}|^{2}}\hat{n}\cdot\hat{r}R_{i}\Delta
Rd\phi
\end{equation}

Finally, we add the contributions of all the inner annuli together
to get the flux $E_{0}$,
\begin{equation}
E_{0}=\Sigma_{i}E_{in}^{i}
\end{equation}
and the equivalent $\mu_{0}$ ( cosine of the equivalent incident
angle $\overline{\beta}$ ) :
\begin{equation}
\mu_{0}=cos\overline{\beta}=\frac{\Sigma_{i}(E_{in}^{i}\mu_{i})}{\Sigma_{i}E_{in}^{i}}
\end{equation}

\section{Emergent flux and image}
To derive the total flux and the image detected by the observer on
earth, the flared surface needs to be projected onto the plane of
sky . For a tiny patch of the flared surface $dS_{A}$, with area
vector $\vec{dS_{A}}$ normal to the surface, the projection area to
the plane which is normal to our line of sight is
$dS_{A}^{\bot}=\vec{dS_{A}}\cdot \hat{l}$, where $\hat{l}$ is the
unit vector pointing to the observer along our line of sight. The
angle between $\hat{l}$ and $\vec{dS_{A}}$
 is $\varphi$=cos$^{-1}$($\hat{l}\cdot \vec{dS_{A}}$/$|\vec{dS_{A}}|$).
The flux we observed from $dS_{A}$ can be derived by multiplying the
intensity at direction $\varphi$, $I_{\varphi}$, and the solid angle
extended by $dS_{A}$ towards us, $F_{A}=I_{\varphi}
dS_{A}^{\bot}/d^{2}$, $d$ is the distance between the object and the
observer. The addition over all the surface, $\int I_{\varphi}
dS_{A}^{\bot}/d^{2}$, is the total flux. The image of the flared
surface could also be derived in this way.

\begin{figure}
\epsscale{.80} \plotone{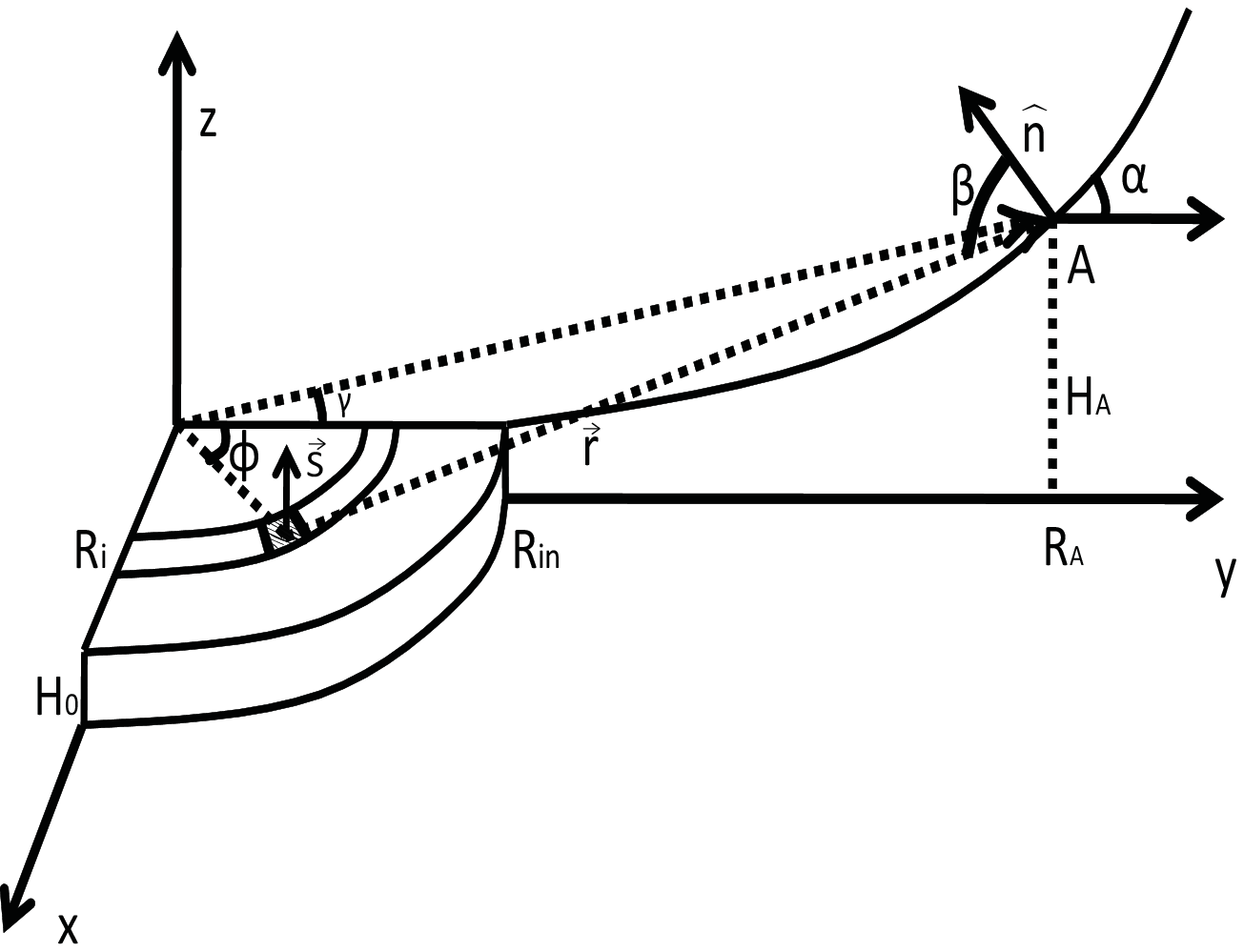} \caption{The geometry used to
calculate the irradiation and the emergent flux. The inner disk has
the radius R$_{in}$ and a height H$_{0}$. The absorption height of
the outer disk at R$_{A}$ is H$_{A}$, and the normal to the flared
surface is $\hat{n}$. } \label{fig:1}
\end{figure}

\begin{figure}
\epsscale{.80} \plotone{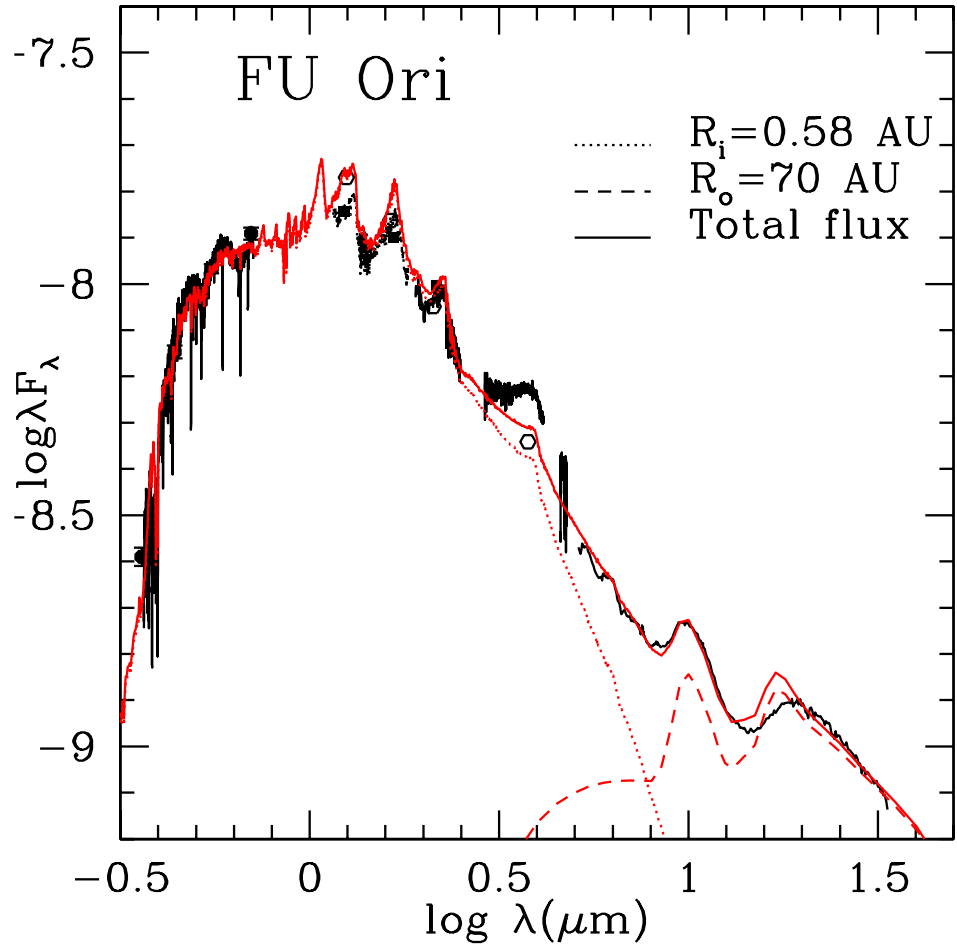} \caption{SED of FU Ori and the
adopted model. Dark curves and dots are observed spectra and
photometry. The light dotted curve is the SED of the inner hot disk
with outer radius R$_{i}$=0.58 AU, while the light dashed curve is
the SED of the flared outer disk with outer radius R$_{o}$=70 AU.
The light solid curve is the total SED with both of the inner and
outer disk.} \label{fig:fuorispe}
\end{figure}

\begin{figure}
\epsscale{.80} \plotone{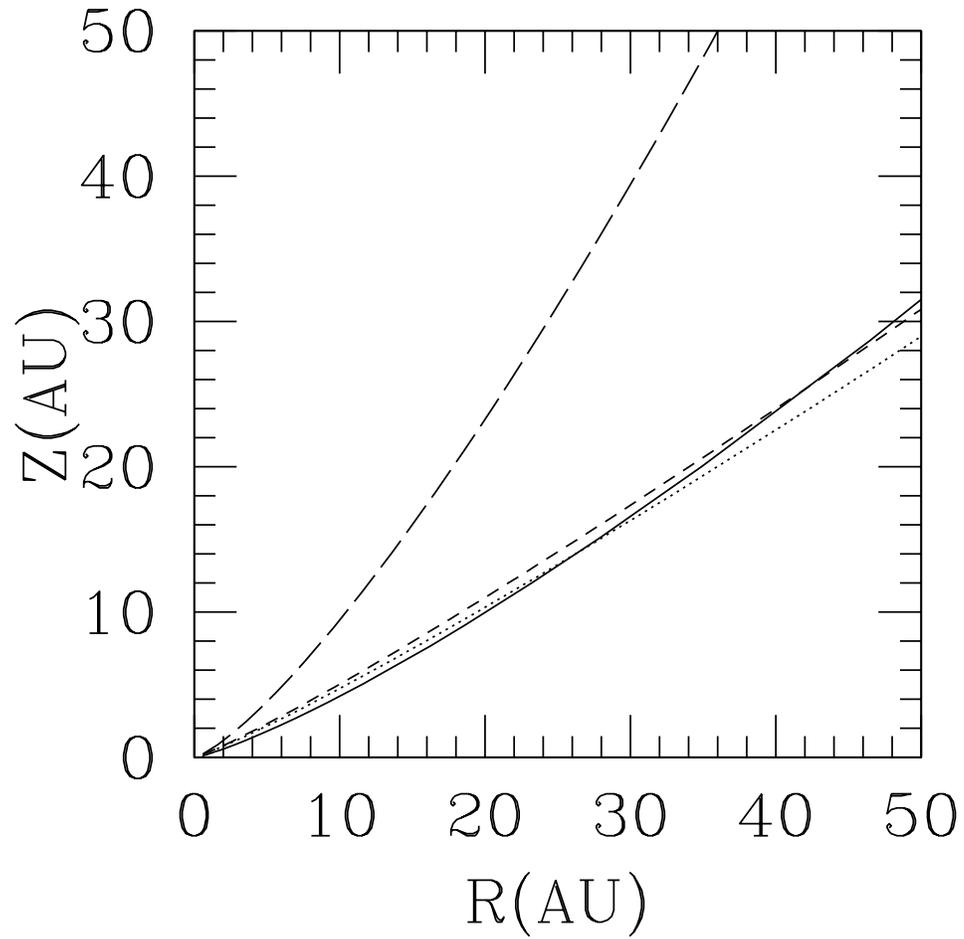} \caption{
The absorption height of the surface of the outer disk for FU
Ori(dotted line), BBW 76 (dashed line) and V1515 Cyg (long dashed
line) as described in Eq. (\ref{eq:surf}). The solid line
corresponds to three scale height of the FU Ori disk (see text).}
\label{fig:fuorisurf}
\end{figure}

\begin{figure}
\epsscale{.80}\plotone{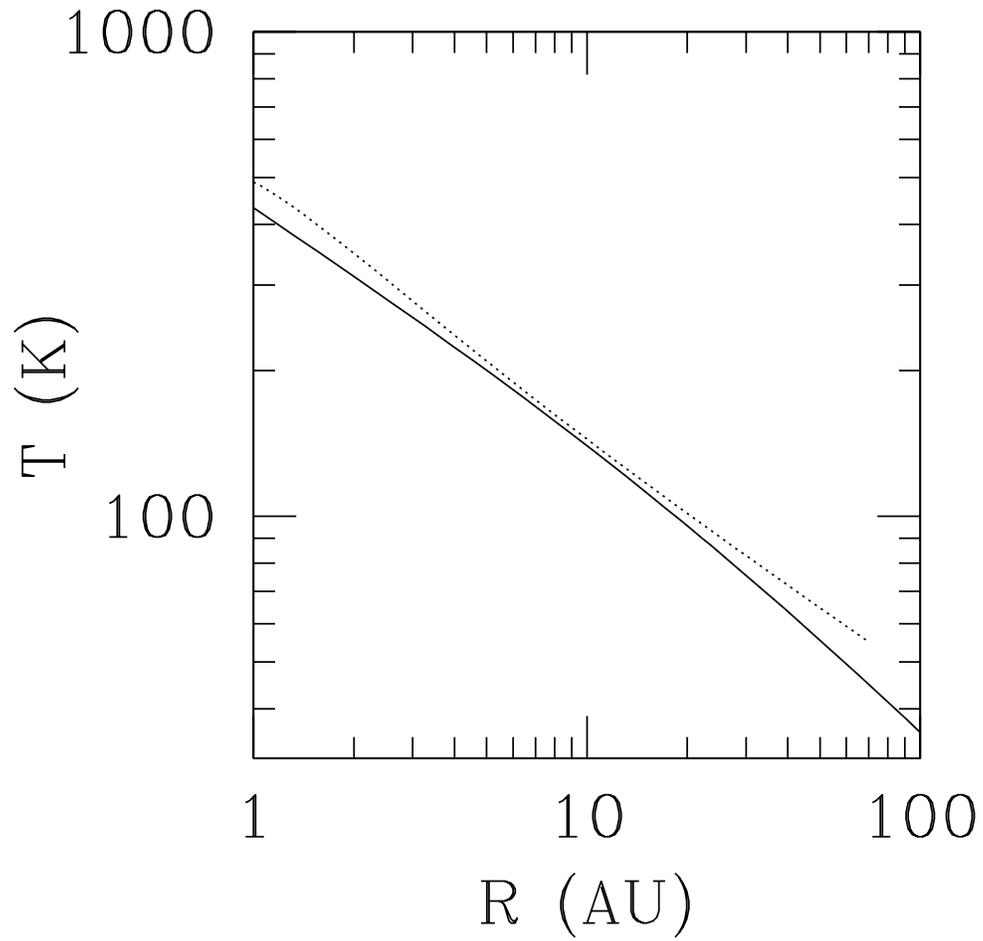}\caption{The effective temperature of
V1515 Cyg (solid line) and FU Ori (dotted line) with respect to the
radius. }\label{fig:surftemp}
\end{figure}

\begin{figure}
\epsscale{.80} \plotone{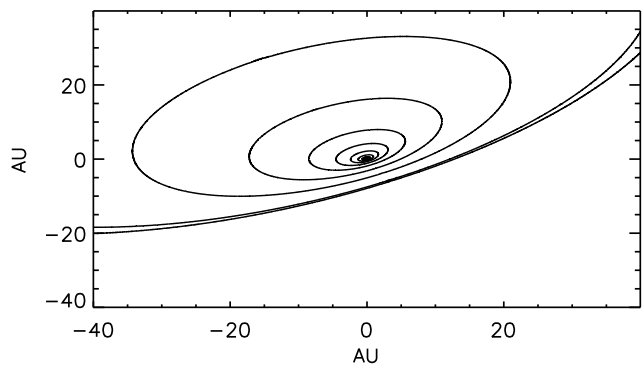} \caption{The H band image
of the FU Ori model with -30$^{o}$ position angle.
Each contour corresponds to a decrease a factor of 4 in intensity,
starting at the innermost contour. } \label{fig:image}
\end{figure}

\begin{figure}
\epsscale{.80} \plotone{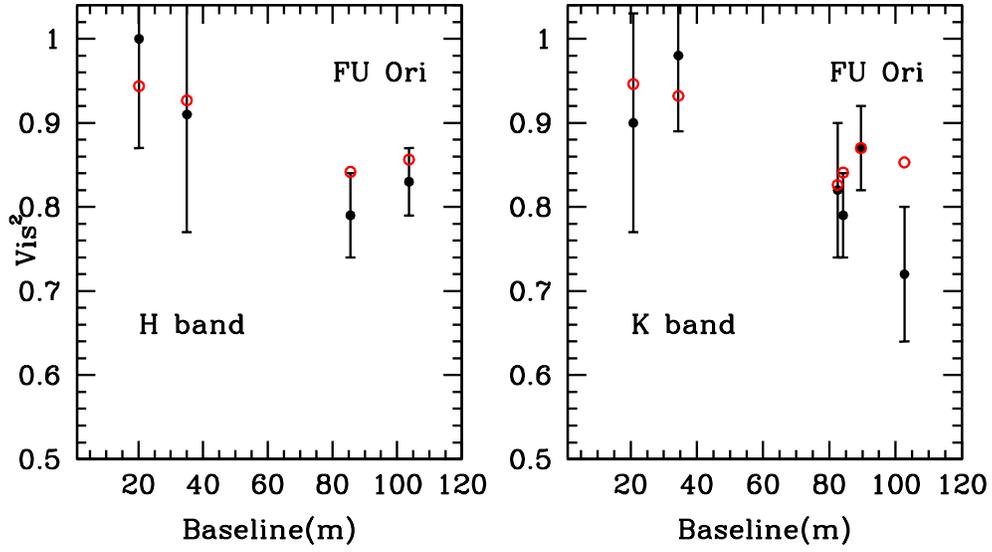} \caption{Visibility square-baseline
at H and K band. Solid dots with error bars are the visibilities
from \cite{malbet05} and the dots are the synthetic visibilities
predicted by our model.} \label{fig:HKvis}
\end{figure}

\begin{figure}
\epsscale{.80} \plotone{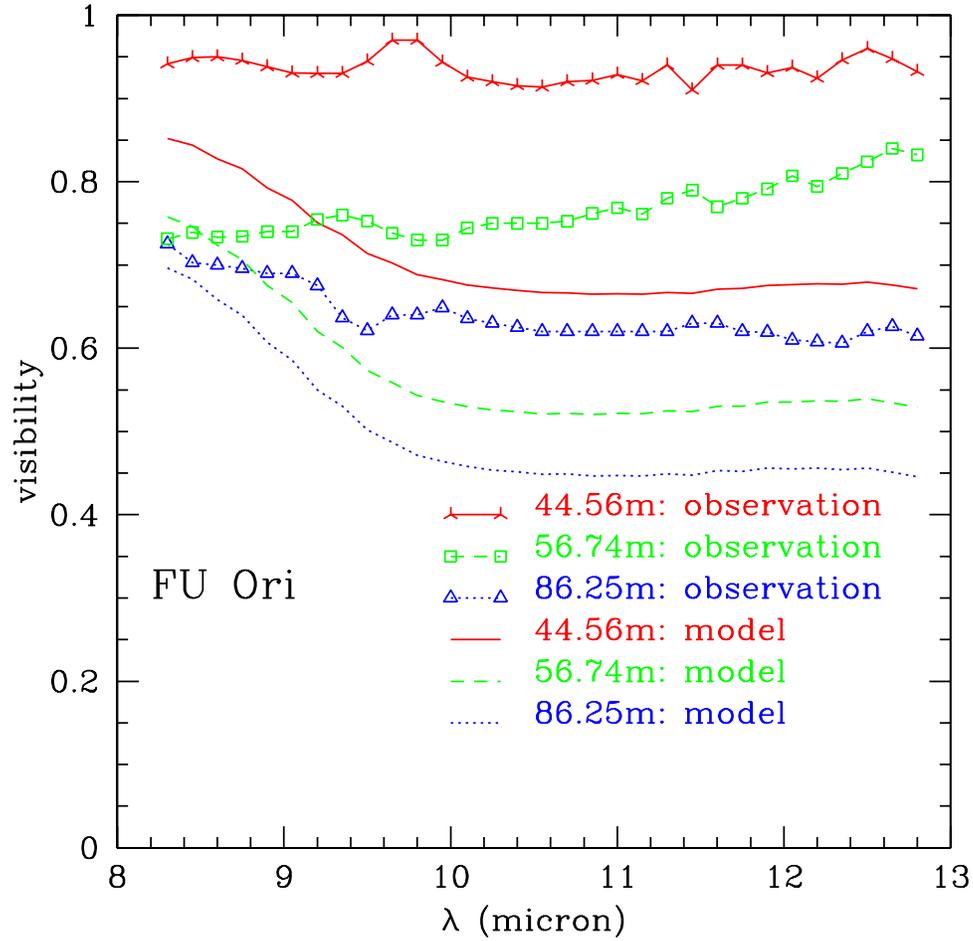} \caption{Visibility-wavelength at
three baselines (44.56 m, 56.74 m, 86.25 m). Observations are from
\cite{quanz2006}. The synthetic visibilities are calculated by
assuming -30$^{o}$ postition angle. } \label{fig:midi}
\end{figure}

\begin{figure}
\epsscale{.80} \plotone{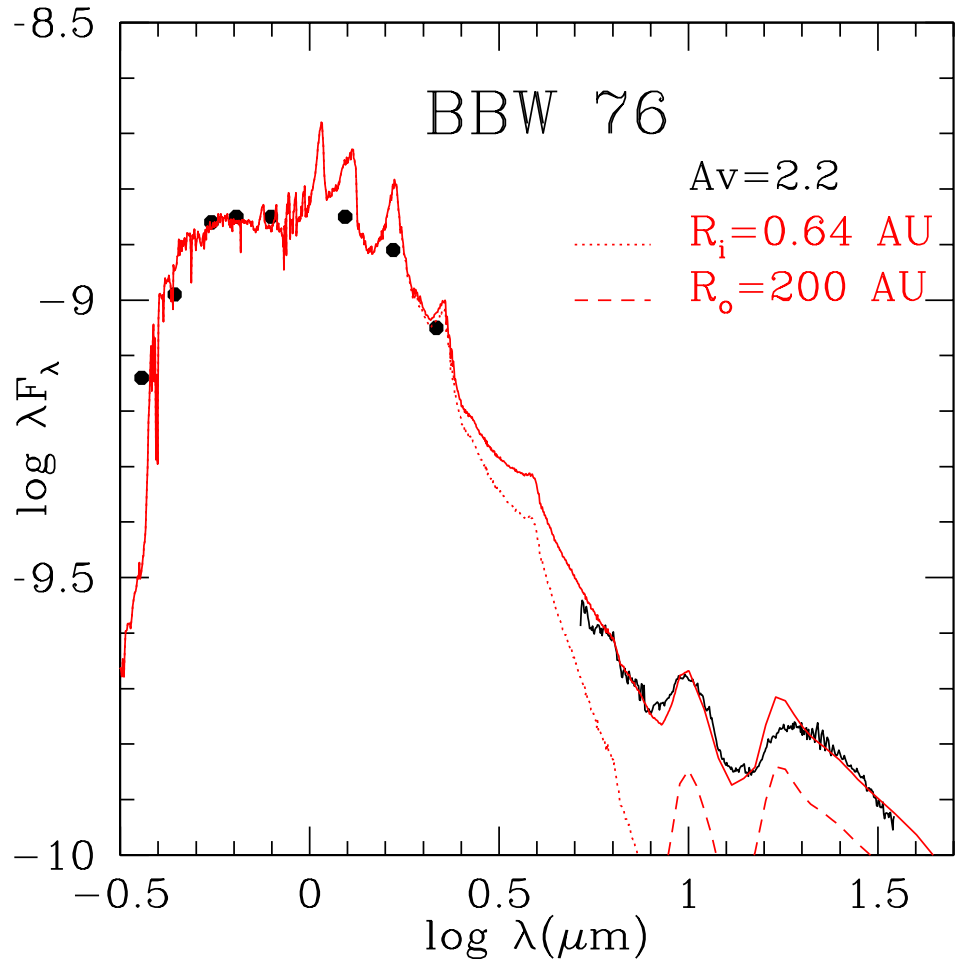} \caption{SED of BBW 76 and adopted
model. Symbols are as in Fig. \ref{fig:fuorispe}.}
\label{fig:BBWspe}
\end{figure}

\begin{figure}
\epsscale{.80} \plotone{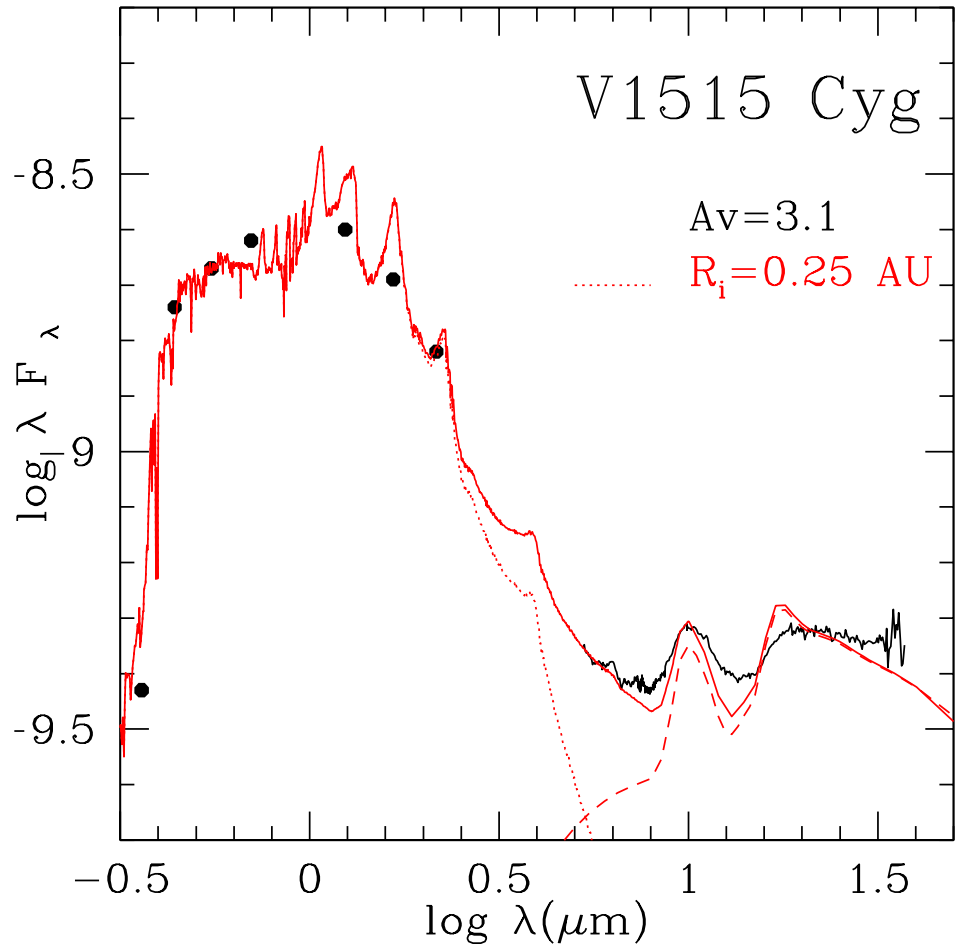} \caption{SED of V1515 Cyg and
adopted model. Symbols are as in Fig. \ref{fig:fuorispe}.}
\label{fig:v1515spe}
\end{figure}

\clearpage
\begin{table}
\begin{center}
\caption{Adopted dust composition \label{tab1}}
\begin{tabular}{cll}
\tableline\tableline
Ingredient  &  $\zeta$\tablenotemark{a} & T$_{sub}$ (K)\tablenotemark{b} \\
\tableline
 Mg$_{0.8}$Fe$_{1.2}$SiO$_{4}$ (Olivine) & 0.0017 & 1460 \\
 Mg$_{0.8}$Fe$_{0.2}$SiO$_{3}$ (Pyroxene)& 0.0017 & 1460 \\
 Graphite & 0.0041 & 734 \\
 Water ice & 0.0056 & 146 \\
\tableline
\end{tabular}
\tablenotetext{a}{Dust-to-gas mass ratio of the particular
ingredient} \tablenotetext{b}{Sublimation temperature at the gas
density $\rho$=10$^{-10}$g/cm$^{3}$}
\end{center}
\end{table}
\clearpage

\begin{table}
\begin{center}
\caption{Parameters for best fit models \label{tab2}}
\begin{tabular}{cccccccccc}
\tableline\tableline
Object  & A$_{V}$ & Inclination &  $M\dot{M}$ & $R_{*}$  & $L_{d}$ & $R_{in}$ \tablenotemark{a}  & $H_{0}$ &  $\gamma$  & $R_{o}$ \tablenotemark{b}\\
& & angle &10$^{-5}$M$_{\odot}^{2}$yr$^{-1}$& R$_{\odot}$&
L$_{\odot}$&AU&R$_{\odot}$ &&AU
 \\
\tableline
 FU Ori & 1.5 & 55 & 7.4 & 5 & 232 & 0.58 & 41.5 & 1.125 & 70 :\\
 BBW 76 & 2.2 & 50 & 8.1 & 4.6 & 277 & 0.64 & 49.22 & 1.125 & 200 :\\
 V1515 Cyg & 3.2 & 0 & 1.3 & 2.8 & 73 & 0.25 & 16.8 & 1.3 & 10$^{4}$ :\\
\tableline
\end{tabular}
\tablenotetext{a}{R$_{in}$ and H$_{0}$ are described in Eq.
\ref{eq:surf}} \tablenotetext{b}{The outer radii are poorly
constrained by the data.}
\end{center}
\end{table}

\end{document}